# MODEL CHECKING FOR MULTI-AGENT SYSTEMS MODELED BY EPISTEMIC PROCESS CALCULUS

Qixian Yu[1], Zining Cao[1,2,3*], Zong Hui[1,4] and Yuan Zhou[1]

[1]College of Computer Science and Technology, Nanjing University of Aeronautics and Astronautics, Nanjing 211106, P. R. China
[2]Ministry Key Laboratory for Safety-Critical Software Development and Verification, Nanjing 211106, P. R. China.
[3]Collaborative Innovation Center of Novel Software Technology and Industrialization, Nanjing 210023, P. R. China
[4]Faculty of Computer and Software Engineering, Huaiyin Institute Of Technology, Huaian 223001, P. R. China
[*]Corresponding author. Email:caozn@nuaa.edu.cn

## ABSTRACT

*This paper presents a comprehensive framework for modeling and verifying multi-agent systems. The paper introduce an Epistemic Process Calculus for multi-agent systems, which formalizes the syntax and semantics to capture the essential features of agent behavior interactions and epistemic states. Building upon this calculus, we propose ATLE, an extension of Alternating-time Temporal Logic incorporating epistemic operators to express complex properties related to agent epistemic state. To verify ATLE specifications, this paper presents a model checking algorithm that systematically explores the state space of a multi-agent system and evaluates the satisfaction of the specified properties. Finally, a case study is given to demonstrate the method.*

## KEYWORDS

*Multi-agent System, Epistemic Logic, Value-Passing CCS, ATL, Model Checking*

## 1. INTRODUCTION

A multi-agent system (MAS) [1] refers to a system composed of multiple autonomous agents that interact, collaborate, or compete with one another to achieve specific tasks or objectives. These agents, which can be software entities, robots, or even humans, operate in dynamic environments where their actions are often influenced by their knowledge, beliefs, and the behaviors of other agents. In recent years, the rapid advancement of artificial intelligence (AI) has led to an explosion of applications leveraging MAS, ranging from autonomous robotics and smart grids to collaborative systems for multiple unmanned aerial vehicles (UAVs) [2]. The ability of MAS to handle complex, distributed tasks has made them increasingly attractive in industrial and societal applications, where they are being adopted to enhance efficiency, scalability, and adaptability in problem-solving scenarios.

One of the key challenges in designing and analyzing MAS lies in effectively modeling the concurrent interactions and cognitive states of agents. Traditional formal methods, such as automata, often fall short in capturing the intricate interplay between agents' behaviors and their epistemic states. In contrast, process calculi provide a powerful framework for modeling concurrent systems due to their ability to handle communication, synchronization, and compositional modeling. Process calculi, such as the Calculus of Communicating Systems (CCS) and $\pi$-calculus, offer expressive mechanisms for describing the dynamic interactions and message-passing behaviors inherent in MAS. This paper builds on these foundations by adopting value-passing CCS[3], which extends traditional process calculi to include explicit data parameters in communication actions. This enhancement allows for more flexible and natural modeling of information exchange and cognitive states among agents, making it particularly suitable for MAS.

To further address the cognitive aspects of agents, this paper introduces the Epistemic Process Calculus (EPC), a novel calculus that integrates value-passing CCS with epistemic logic. EPC enables the modeling of not only the behavioral interactions of agents but also their knowledge and beliefs, which are critical for decision-making in MAS. Once the system is modeled using EPC, it becomes necessary to specify and verify its properties. For this purpose, we extend Alternating-time Temporal Logic (ATL) with epistemic operators, resulting in a new logic called ATLE (Alternating-time Temporal Logic with Epistemic operators). ATLE allows for the expression of complex properties related to agents' knowledge, strategic interactions, and temporal behaviors. Finally, to verify these properties, we propose a model checking algorithm specifically designed for ATLE, which systematically explores the state space of the MAS and evaluates the satisfaction of the specified properties. Through this integrated approach, our framework provides a comprehensive solution for modeling, specifying, and verifying multi-agent systems, ensuring their correctness and reliability in dynamic and uncertain environments.The primary contributions of this paper can be encapsulated as follows:

(1) The paper introduces a novel process calculus called epistemic process calculus(EPC) tailored for multi-agent systems, which it delineates with formal precision.

(2) It proposes a novel logic ATLE for EPC and devises a corresponding model checking algorithm to facilitate its evaluation.

(3) A case study has been employed to demonstrate the feasibility of employing EPC modeling for representing the properties of a system using ATLE, followed by the verification of the model through model checking algorithms.

The paper is structured as follows: Section 2 reviews model checking advancements for multi-agent systems. Section 3 introduces the syntax and semantics of EPC. Section 4 details a model checking algorithm for ATLE. Section 5 demonstrates how to implement modeling and model detection algorithms at the code level. Section 6 presents a case study. Section 7 concludes with a summary and directions for future research.

## 2. RELATED WORK

Model checking is a critical technique for safety verification. The process of model checking is divided into three stages: modeling, specification, and verification. In the modeling phase of multi-agent systems (MAS), common approaches include automata[4], Petri nets[5], and process calculi[6]. However, formal modeling methods other than process calculi often fall short in terms of synchronization, concurrency, and compositional modeling, which are core requirements for MAS. Existing methods based on process calculi for modeling MAS generally have some limitations. Firstly, many methods (such as CAML[7], EMMAS[8], and extensions based on SALMA[9]) are overly complex at the semantic level, increasing the difficulty of learning and implementation, as well as raising the barrier to practical application. Secondly, these methods lack adaptability in dynamic environments, making it difficult to effectively handle dynamic changes in agent behavior and adjustments in interaction patterns. Additionally, these methods often lack sufficient consideration and support for the cognitive aspects of agents in MAS.

Value-passing CCS (Calculus of Communicating Systems)[10], compared to traditional CCS, retains the simplicity of algebraic semantics and rigorous derivation mechanisms while introducing explicit data parameters in communication actions. This makes the information exchange and cognitive state modeling among agents more flexible and natural. However, the academic community has rarely extended value-passing CCS for MAS modeling, primarily because value-passing CCS has limited capabilities in dynamic communication topology and mobility support, making it difficult to meet the flexible and dynamic interaction requirements of MAS. Furthermore, value-passing CCS lacks direct support for cognitive mechanisms and has insufficient composability and extensibility, limiting its application in complex MAS scenarios. In contrast, $\pi$ − calculus [11], with its dynamic name migration and high expressiveness, is more suitable for describing the dynamic interactions and concurrent behaviors of MAS. Value-passing CCS is typically applied in distributed systems [12], network protocols[13], programming language theory, and security analysis, where it is used to model and verify communication protocols and distributed systems.

This paper uses value-passing CCS as a foundation and combines it with dynamic epistemic logic to create the Epistemic Process Calculus (EPC). EPC enhances communication flexibility and introduces cognitive modeling, addressing the limitations of value-passing CCS in MAS. However, manual modeling remains challenging, and future work should focus on improving EPC-based modeling and verification through interdisciplinary approaches and tool development.

Extending Alternating-time Temporal Logic (ATL)[14-15] to describe system properties is a natural choice for EPC. ATL's ability to describe strategic interactions aligns well with EPC's dynamic nature, and its semantic framework can easily incorporate epistemic operators to model agent cognition. Together, EPC and ATLE logic enable a complete pipeline from modeling and specification to verification.

## 3. EPISTEMIC PROCESS CALCULUS

The Epistemic Process Calculus (EPC) is employed to model multi-agent systems (MAS) due to its ability to seamlessly integrate concurrency, dynamic communication, and cognitive reasoning. EPC extends traditional process calculi by incorporating epistemic logic, enabling the explicit representation of agents' knowledge and beliefs, which is crucial for capturing the complex interactions and decision-making processes in MAS.

### 3.1. Syntax

The syntax is given for a set of agents Ag. Processes represent one or more executions of an individual agent, and the smallest unit of process execution is the action, which is an atomic operation representing the basic capability of the process. Name is a basic concept that identifies the communication channel or the subject of the communication. Names have special meanings in EPC; they can denote a channel, a process, or another communicating entity. Define $\mathbb{D}$ as the data set, and $\mathbb{K}$ as the knowledge set, They are the fundamental units for processes to send and receive, and they are all values. The set of values is denoted as Values, thus $\mathbb{D} \cup \mathbb{K}$ = Values. Actions represent the roles and behaviors of agents within communication events. Let Var be the set of variables, Ag is the set of agents of multi-agent system.

Definition 3.1: (Action). The action represents the role and behavior of the agent in a communication event. The action act is specified as follows：

$$act \coloneqq \bar{a} < t > |a(x)|\tau, \text{ where } x \in Var, t \in \mathbb{D} \cup \mathbb{K}, \text{ and } a, \tau \in Name$$

Here, x represents a variable in the data that receives content through channel a, the specifics of which are uncertain, hence it is considered a variable, whereas t denotes a definite piece of data or knowledge. $\bar{a} < t >$ represents the capability of a process to send data or knowledge after execute the action. a(x) denotes the ability of a process to receive a variable after executing an action. τ is an action internal to a process, an internal action. Acts is a non-empty set of actions, its represents all action that process can execute.

Definition 3.2: (Process). Utilizing the syntactic rules of process caculus, we construct process expressions that delineate the behavior of agents based on their actions and modes of interaction. The specific process P is defined by the BNF formalism as follows::

$$P := 0 | act.P | (va)P | (P|P) | P + P | C$$

In the above, a ∈ Name, 0 means that there is no action prefix process, i.e. "dead" process, the process does not perform any operation; act.P indicates that process P is guarded by the action prefix and P must be executed only after the action is finished; C denotes a constant, $C \stackrel{def}{=} P$, Where P is a process expression, which can be defined recursively.

Definition 3.3: (Labeled Process). A labeled process includes a global environment and several agents, where agents are denoted by uppercase letters (except for M and P). The behavior of each agent is controlled by one or more processes P. The BNF of the labeled process M is as follows:

$$M := (\{P\}_i) | (M|M) | (va)M$$

Here a ∈ Name, i ∈ Ag, $\{P\}_i$ denotes a system composed of unique agent i and the behavior of i is controlled by the process P. M|M represents a composite on the system, indicating that the system consists of multiple subsystems The language has two constraint constructions: the restriction name (va)M and the input a(x).

In EPC, there is a constraint constructs: the restriction(va)P. The sets of bound names and free names in process P are denoted by bn(P) and fn(P), respectively. Processes that differ only in their bound names are called α-equivalent, denoted by $\equiv_\alpha$. This paper does not distinguish between processes that are α-equivalent. A substitution is a partial mapping from Name to Name, denoted by σ. Substitution is a postfix operator with higher precedence than any other operator in the language.

Definition 3.4: (Structural Congruence). The smallest relation ≡ contains the relationship $\equiv_\alpha$, ≡ satisfying the following axioms:

$$P|Q \equiv Q|P \quad P|0 \equiv P$$
$$P_1|(P_2|P_3) \equiv (P_1|P_2)|P_3$$
$$P_1 + (P_2 + P_3) \equiv (P_1 + P_2) + P_3 \quad P + 0 \equiv P$$
$$(va)(vb)P \equiv (vb)(va)P$$
$$P|(va)P \equiv (va)(P|Q) \text{ if } a \notin fn(P)$$

### 3.2. Semantics

In EPC, Structural Operational Semantics (SOS) rules define process behaviors and transitions through formal inference rules. They specify state changes precisely and underpin the analysis, verification, and model checking of concurrent systems, refer to Table 1 and Table 2.

In Table 2, there are numerous actions accompanied by labels, which in the form of labeled actions. Labeled actions are defined as all actions that can be driven by agents with corresponding labels. Let the set of labels is denoted as LabelActs = $\{\{\alpha\}_i, \{\tau\}_{i,j} | \alpha = a < t >$ or $\bar{a} < t >$, $i, j \in Ag, t \in Values\}$.

Table 1. SOS Rules of Process

$$(\text{Pref}_1) \ \frac{}{\alpha.P \xrightarrow{\alpha} P}, \alpha \in \{\tau, \bar{a}<t>\}$$

$$(\text{Pref}_2) \ \frac{}{a(x).P \xrightarrow{a<t>} P\{t/x\}}$$

$$(\text{Sum}_1) \ \frac{P \xrightarrow{\alpha} P'}{(P+Q) \xrightarrow{\alpha} P'} \quad (\text{Sum}_2) \ \frac{Q \xrightarrow{\alpha} Q'}{(P+Q) \xrightarrow{\alpha} Q'}$$

$$(\text{Par}_1) \ \frac{P \xrightarrow{\alpha} P'}{P|Q \xrightarrow{\alpha} P'|Q} \quad (\text{Par}_2) \ \frac{Q \xrightarrow{\alpha} Q'}{P|Q \xrightarrow{\alpha} P|Q'}$$

$$(\text{Com}) \ \frac{P \xrightarrow{a<t>} P', Q \xrightarrow{\bar{a}<t>} Q'}{P|Q \xrightarrow{\tau} P'|Q'} \quad (\text{Res}) \ \frac{P \xrightarrow{\alpha} P'}{(va)P \xrightarrow{\alpha} (va)P'} \ \alpha \neq a<t>, \bar{a}<t>$$

$$(\text{Con}) \ \frac{P \xrightarrow{\alpha} P'}{C \xrightarrow{\alpha} P'}, C \stackrel{\text{def}}{=} P$$

Table 2. SOS Rules of Labeled Process

$$(\text{PrefM}) \ \frac{}{\{P\}_i \xrightarrow{\{\alpha\}_i} \{P'\}_i}, \alpha \in \{\tau, \bar{a}<t>, a<t>\}$$

$$(\text{ParM}_1) \ \frac{M \xrightarrow{\{\alpha\}_i} M'}{M|N \xrightarrow{\{\alpha\}_i} M'|N} \quad (\text{ParM}_2) \ \frac{N \xrightarrow{\{\alpha\}_i} N'}{M|N \xrightarrow{\{\alpha\}_i} M|N'}$$

$$(\text{ComM}) \ \frac{M \xrightarrow{\{a<t>)\}_i} M', N \xrightarrow{\{\bar{a}<t>\}_j} N'}{M|N \xrightarrow{\{\tau\}_{i,j}} M'|N'}$$

$$(\text{ResM}) \ \frac{M \xrightarrow{\{\alpha\}_i} M'}{(va)M \xrightarrow{\{\alpha\}_i} (va)M'}, \alpha \neq a<t>, \bar{a}<t>$$

This paper employs a set of symbols and functions to denote certain specialized semantics. $S=S_0 \times S_1 .... \times S_n$ is a set of non-empty global state, $S_i$ represents the non-empty sate set of agent i. MS is a set of non-empty labeled process set. SMS=S×MS is the non-empty set of the tuple of state and labeled process. AP is a set of atomic propositions. T=S×AP→Bool is an abstract function that maps the propositions under the global state S. $K \subseteq S \times \text{LabelActs} \times S$ is a transition relation representing the state evolution of the system.

Multi-agent systems are established based on epistemic states, and their evolution is dependent on changes in these epistemic states. In this paper, we denote the configuration of a system as (s, M), which includes the state and the processes, while $(s, M) \xrightarrow{\text{act}} (s', M')$ represents the transition of the system from one configuration to the next configuration. The definition of this relationship is as follows:

$$\text{If } M \xrightarrow{act} M', \ act \in \text{LabelActs and } (s, act, s') \in K, \ (s, M) \xrightarrow{act} (s', M')$$

This indicates that under labeled process m, the state s evolves in to labeled process m' through the execution of action act, and the system changes from m to m'. Table 1 pertains to the SOS rules for processes, while Table 2 pertains to the SOS rules for labeled processes. Moreover, the epistemic state of each agent can be represented by an abstract function: $h_i(s) = es_i$. The function $h(s) = (h_0(s), h_1(s), \ldots, h_i(s)), i \in Ag$ is tasked with abstracting the epistemic state from the global state, or with comprehensively abstracting the epistemic state pertinent to agents from the entire state context.

In multi-agent systems, strategies are essential for modeling how agents make decisions and interact with each other to achieve specific goals. A strategy defines the actions an agent or a coalition of agents can take in different states, guiding the system's evolution over time. By incorporating strategies into the model, we can formally verify whether certain properties, such as safety or liveness, hold under specific agent behaviors. This is particularly important in dynamic environments where agents must adapt their actions based on their knowledge and the state of the system. Strategies enable the analysis of cooperative or competitive behaviors among agents, ensuring that the system meets its desired specifications.

We define the notion of strategies. Consider a labeled process M. For agents $A \in 2^{Ag}, U \subseteq S$, a partial strategy $f_A^U$ maps the next action that can be executed asynchronously by the set of agents A. When an agent not in A performs an action, it is arbitrary; therefore, when executing a partial strategy $f_A^U$ in some states, it is the agents not in A who perform certain actions. $F_A(f_A^U)$ maps the actions taken under the strategy $f_A^U$ across all states. Let $\text{Acts}_A = \{\{\alpha\}_i, \{\tau\}_{i,j} | i, j \in A, \alpha \in \{a < t >, \bar{a} < t > \}$, $f_A^U$ and $F_A(f_A^U)$ can be defined as follows:

$$f_A^U \in \{U \to \text{Acts}_A, U \subseteq S\}$$

$$F_A(f_A^U) = \{f_A | U \subseteq S, f_A(s) = f_A^U(s), \text{if } s \in U; f_A(s) \in \text{Acts}_{Ag-A}, \text{if } s \notin U\}$$

If labeled process is M, the $\text{out}(f_A, (s, M))$ represents all configuration path formed by iteratively executing the strategies in $f_A$ starting from (s, M). A path starts from $(s_0, M_0)$ after a series of actions: $\pi = (s_0, M_0)(s_1, M_1)\ldots(s_k, M_k)$, $\pi \in \text{out}(f_A, (s_0, M_0))$, this sequence satisfies the following condition: $f_A(s_i) = act_i$ and $(s_i, M_i) \xrightarrow{act_i} (s_{i+1}, M_{i+1}), 0 \leq i < k$. $\pi[i]$ represents the i-th element in this sequence, $\pi[1] = (s_1, M_1)$.

Next, we will use a simple case to illustrate the use of strategies. Suppose there are three agents $Ag = \{1,2,3\}$ using EPC modeling as follows:

$$P = \bar{c} < t_1 > . \bar{a} < t_2 > . 0 | \bar{d} < t_3 > . 0, \ Q = c(x).a(x).\bar{b} < t_2 > . 0, \ R = d(x).0$$

$$M_0 = \{P\}_1 | \{Q\}_2 | \{R\}_3$$

Here, only one transition is presented for demonstration, while the entire transitions can be observed in the Fig.1.

$$M_0 \xrightarrow{\{\tau\}_{1,2}} M_1 = \{\bar{a} < t_2 > . \bar{b} < t_2 > . 0 | \bar{d} < t_3 > . 0\}_1 | \{a(x).0\}_2 | | \{d(x).0\}_3$$

$$\xrightarrow{\{\tau\}_{1,2}} M_2 = \{\bar{b}<t_2>.0|\bar{d}<t_3>.0\}_1|\{0\}_2|\{d(x).0\}_3$$

$$\xrightarrow{\{\bar{b}<t_2>\}_2} M_3 = \{\bar{d}<t_3>.0\}_1|\{0\}_2|\{d(x).0\}_3 \xrightarrow{\{\tau\}_{1,3}} M_4 = \{0\}_1|\{0\}_2||\{.0\}_3$$

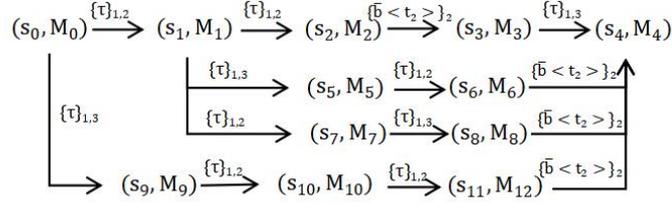

Fig.1. The Diagram of Systemic Transition

Define $U = \{s_0, s_1, s_2, s_5, s_6, s_8, s_9, s_{10}\}$, $A = \{1,2\}$, $f_A^U = \{\{\tau\}_{1,2}, \{\bar{b}<t_2>\}_2\}$. In this context, U denotes a collection of states, indicating that agents within set A are capable of performing actions in accordance with policy $f_A^U$ when in states that are included in U. Conversely, for states that are not part of U, actions are selected by agents outside of set A. Then according to the strategy $out(f_A, s)$, a path is generated as follows: $\pi_0 = (s_0, M_0)(s_1, M_1)(s_2, M_2)$. Only this sequence satisfies the $F_A(f_A^U)$, whereas others, such as $(s_0, M_0)\xrightarrow{\{\tau\}_{1,3}}(s_5, M_5)$, $s_0 \in U$, $\{\tau\}_{1,3} \notin f_A^U(s_0)$.

## 4. MODEL CHECKING

Alternating-time Temporal Logic with Epistemic (ATLE) is used for model checking in multi-agent systems because it combines the strengths of temporal logic and epistemic logic, enabling the specification and verification of complex properties related to both time and knowledge. ATLE extends Alternating-time Temporal Logic (ATL)[14] by incorporating epistemic operators, allowing us to express not only temporal properties (e.g., "eventually" or "always") but also knowledge-based properties (e.g., "an agent knows that a condition holds"). This is crucial in multi-agent systems, where agents' actions often depend on their knowledge of the environment and other agents. By using ATLE, we can formally verify properties such as whether a coalition of agents can achieve a goal based on their shared knowledge, or whether certain safety or liveness conditions hold under specific strategies.

### 4.1. ATLE

Definition 4.1: (Syntax of ATLE ). In the set of atomic propositions AP, let the number of agents be denoted by Ag = 1,...,n. The ATLE formula ϕ indicates that a certain property ϕ holds under A:

$$\phi ::= ap \,|\, \neg\phi \,|\, \phi_1 \vee \phi_2 \,|\, K_i(\phi) \,|\, E_A(\phi) \,|\, C_A(\phi) \,|\, D_A(\phi) \,|$$

$$\langle\langle A \rangle\rangle X(\phi) \,|\, \langle\langle A \rangle\rangle(\phi_1 U \phi_2) \,|\, \langle\langle A \rangle\rangle F(\phi) \,|\, \langle\langle A \rangle\rangle G(\phi)$$

Where $ap \in AP$, $A \subseteq Ag$. $\phi, \phi_1, \phi_2$ are ATLE formulas. Moreover, $K_i(\phi)$ represent the agent i know the ϕ, $D_A\phi$ mean that in A, ϕ is distributed knowledge, $C_A(\phi)$ mean that in A, ϕ is common knowledge, $E_A(\phi)$ mean that in A, every agent knows ϕ. Respectively, distributed knowledge is about the combined potential knowledge of a group, while common knowledge requires that everyone in the group knows a fact and is aware that everyone else knows it too...,recursively, leading to an infinite regression of mutual knowledge.

The ATLE formulas $\langle\langle A\rangle\rangle X(\phi)$, $\langle\langle A\rangle\rangle(\phi_1 U \phi_2)$, $\langle\langle A\rangle\rangle F(\phi)$, $\langle\langle A\rangle\rangle G(\phi)$ represent the next, until, eventually, and globally modalities, indicating that there exists a strategy for set A, $\phi$ is true at the next state, there exists a strategy for set A, $\phi_1$ remains true until $\phi_2$, there exists a strategy for set A, $\phi$ eventually becomes true, and there exists a strategy for set A, $\phi$ is always true, respectively.

Definition 4.2: (Semantics of ATLE). Given a labeled process M, a state s and ATLE formula $\phi$, the satisfaction of $\phi$ is denoted as $(s, M) \vDash \phi$, and satisfies the following recursive condition:

- $(s, M) \vDash ap$, ap is atomic proposition iff $T(s, ap) = true$;

- $(s, M) \vDash \neg\phi$ iff $(s, M) \nvDash \phi$;

- $(s, M) \vDash \phi_1 \vee \phi_2$ iff $(s, M) \vDash \phi_1$ or $(s, M) \nvDash \phi_2$;

- $(s, M) \vDash \langle\langle A\rangle\rangle X(\phi)$ iff there exists a set $U \subseteq S$ and a strategy $f_A^U$ for any $f_A \in F_A(f_A^U)$, such that for any paths $\pi \in out(f_A, s)$, we have $\pi[1] \vDash \phi$;

- $(s, M) \vDash \langle\langle A\rangle\rangle G(\phi)$ iff there exists a set $U \subseteq S$ and a strategy $f_A^U$, for any $f_A \in F_A(f_A^U)$, $\pi \in out(f_A, s)$, such that for any paths for all $i \geq 0$, we have $\pi[i] \vDash \phi$;

- $(s, M) \vDash \langle\langle A\rangle\rangle F(\phi)$ iff there exists a set $U \subseteq S$ and a strategy $f_A^U$, for some $f_A \in F_A(f_A^U)$, $\pi \in out(f_A, s)$, such that for any paths for some $0 \leq i < \infty$, we have $\pi[i] \vDash \phi$;

- $(s, M) \vDash \langle\langle A\rangle\rangle(\phi_1 U \phi_2)$ iff there exists a set $U \subseteq S$ and a strategy $f_A^U$, for some $f_A \in F_A(f_A^U)$, such that for any paths $\pi \in out(f_A, s)$, for some $i \geq 0$, we have $\pi[i] \vDash \phi_1$, and for any $0 \leq j < i$, we have $\pi[j] \vDash \phi_2$;

- $(s, M) \vDash K_i \phi$ iff if $h_i(s) = h_i(s')$, we have $(s', M') \vDash \phi$;

- $(s, M) \vDash E_A(\phi)$ iff for any agent $i \in A$, we have $(s, M) \vDash K_i \phi$;

- $(s, M) \vDash C_A(\phi)$ iff for any agent $i \in A$, $R_i = \{((s, M), (s', M')) | h_i(s) = h_i(s')\}$, $R_i^+$ is $R_i$'s transitive closure, if $((s, M), (s', M')) \in R_i^+$, we have $(s', M') \vDash \phi$;

- $(s, M) \vDash D_A(\phi)$ iff exist agent $i \in A$, we have $(s, M) \vDash K_i(\phi)$;

The axioms and inference rules of epistemic logic are sound under the syntactic framework defined in this paper. According to the research[18], the soundness of epistemic logic is closely tied to the definition of its syntax, ensuring that all inference rules maintain logical consistency and correctness within the given syntactic framework Consequently, the syntactic framework proposed in this paper is capable of supporting epistemic reasoning, thereby ensuring the reliable transmission and sharing of knowledge in multi-agent systems.

### 4.2. Model Checking for ATLE

This section introduces a specific model checking algorithm for ATLE specification formulas. Algorithm 1 determines whether a propositional formula $\phi$ holds in a state s for a labeled process M. The input process M must be specified as a finite-state process; otherwise, the algorithm may fail to terminate when handling infinite-state processes. It processes different logical constructs, such as atomic propositions, negations, dis-junctions, and various modal operators including knowledge, common knowledge, and temporal modalities, etc... Each case is handled through specific checks to evaluate the truth of $\phi$ under different logical rules. The algorithms for the knowledge-related operators $K_i$, $E_A$, $D_A$, and $C_A$ as illustrated in Algorithms 2, 3, 4 and 5 respectively. Algorithms 6, 7, 8 and 9 primarily involve the implementation of path operators, corresponding respectively to the operators $\langle\langle A\rangle\rangle X$ (Next), $\langle\langle A\rangle\rangle G$ (Global), $\langle\langle A\rangle\rangle F$ (Finally) and $\langle\langle A\rangle\rangle U$ (Until).

**Algorithm 1** CHECK(s, M, ϕ)

1: **switch**(ϕ):
2:    **case** atomic proposition: **return** T(s, ϕ) ;
3:    **case** ¬ϕ: **return** CHECK(s, M, ϕ);
4:    **case** ϕ$_1$ ∨ ϕ$_2$:
5:       **return** CHECK(s, M, ϕ$_1$) ∨ CHECK(s, M, ϕ$_2$) ;
6:    **case** K$_i$(ϕ):
7:       **return** CHECKK(s, M, ϕ, i);
8:    **case** E$_A$(ϕ):
9:       **return** CHECKE(s, M, ϕ, A);
10:   **case** D$_A$(ϕ):
11:      **return** CHECKD(s, M, ϕ, A);
12:   **case** C$_A$(ϕ):
13:      **return** CHECKC(s, M, ϕ, A);
14:   **case** ⟨⟨A⟩⟩X(ϕ):
15:      **return** CHECKX (s, M, ϕ, A);
16:   **case** ⟨⟨A⟩⟩G(ϕ):
17:      **return** CHECKG(s, M, ϕ, A);
18:   **case** ⟨⟨A⟩⟩F (ϕ):
19:      **return** CHECKF (s, M, ϕ, A);
20:   **case** ⟨⟨A⟩⟩(ϕ$_1$U ϕ$_2$):
21:      **return** CHECKU (s, M, ϕ$_1$, ϕ$_2$, A);
22:   **default**: throw a formula error exception;
23: **end switch**

---

**Algorithm 2** CHECKK(s, M, ϕ, i)

1: **for all** (s', M') in SMS **do**
2:    **if** h$_i$(s') = h$_i$(s) and CHECK (s, M, ϕ) = false **then**
3:       **return** false;
4:    **end if**
5: **end for**
6: label(s) := label(s) ∪ {K$_i$(ϕ)};
7: **return** true;

---

**Algorithm 3** CHECKE(s, M, ϕ, A)

1: **for all** agent i **in** A **do**
2:    **if** CHECKK (s, M, ϕ, i) = false **then**
3:       **return** false;
4:    **end if**
5: **end for**
6: label(s) := label(s) ∪ {E$_A$(ϕ)};
7: **return** true;

---

**Algorithm 4** CHECKD(s, M, ϕ, A)

1: **for all** agent i **in** A **do**
2:    **if** CHECKK(s, M, ϕ, i) = true **then**
3:      label(s) := label(s) ∪ {D$_A$(ϕ)};
4:      **return** true;
5:    **end if**
6: **end for**
7: **return** false;

**Algorithm 5** CHECKC(s, M, ϕ, A)

1: Initialize $R_A^+$ to store transitive closure of relation $R_A$;
2: **if** CHECKK (s, M, ϕ, A) = false **then**
3:    **return** false;
4: **end if**
5: **for all** (s', M') in SMS **do**
6:    **if** ((s, M),(s', M')) ∈ $R_A^+$ and CHECKK (s , M, ϕ, A) = false **then**
7:      **return** false;
8:    **end if**
9: **end for**
10: label(s) := label(s) ∪ {$C_A(ϕ)$};
11: **return** true;

---

**Algorithm 6** CHECKX (s, M, ϕ, A)

1: **function** CHECKX (s, M, ϕ, A):
2:   **for all** $f_A^U$ **do**
3:     **if** checkX(s,M,ϕ,$f_A^U$) = true **then**
4:       **return** true;
5:     **end if**
6:   **end for**
7:   **return** false;
8: **end function**
9: **function** checkX(s, M, ϕ, $f_A^U$)
10:   **for all** act in $F_A(f_A^U)(s)$ **do**
11:     **if** ((s, M) $\xrightarrow{act}$ (s', M') and CHECK(s', M', ϕ) = false **then**
13:       **return** false;
14:     **end if**
15:   **end for**
16:   label(s) := label(s) ∪ {⟨⟨A⟩⟩X(ϕ)};
17:   **return** true;
18: **end function**

---

**Algorithm 7** CHECKG(s, M, ϕ, A)

1: **function** CHECKG(s, M, ϕ, A)
2:   **for all** $f_A^U$ **do**
3:     SCCs = findSCCS(s, M, $F_A(f_A^U)$);
4:     **if** checkG(SCCs, $f_A^U$, ϕ) = true **do**
5:       **return** true;
6:     **end if**
7:   **end for**
8:   **return** false;
9: **end function**
10: **function** checkG(SCCs , ψ):
11:   **for all** scc **in** SCCs **do**
12:     **for all** (s,M) **in** scc **do**
13:       **if** ϕ ∉ label(s) or CHECK(s, M, ϕ) = false **then**
14:         **return** false;
15:       **end if**
17:     **end for**
18:   **end for**

19:   label(s) := label(s) ∪ {⟨⟨A⟩⟩G(ϕ)};
20:   **return** true;
21:**end function**

---

**Algorithm** 8 CHECKF (s, M, ϕ, A)

1:**function** CHECKG(s, M, ϕ, A)
2:   **for all** $f_A^U$ **do**
3:      SCCs := findSCCS(s, M, $F_A(f_A^U)$);
4:      **if** checkF(SCCs, ϕ) = true **do**
5:         **return** true;
6:      **end if**
7:   **end for**
8:   **return** false;
9:**end function**
10:**function** checkF(SCCs, ϕ):
11:   **for all** sccc **in** SCCs **do**
12:      **for all** (s', M') is (s, M)'s successor **do**
13:         **if** CHECK(s', M', ϕ) = true **then**
14:            label(s) := label(s) ∪ {⟨⟨A⟩⟩Fϕ};
15:            **return** true;
16:         **end if**
17:      **end for**
18:   **end for**
19:   **return** false;
20:**end function**

---

**Algorithm** 9 CHECKU (s, M, ϕ$_1$, ϕ$_2$, A)

1:**function** CHECKU (s, M, ϕ$_1$, ϕ$_2$, A):
2:   **for all** $f_A^U$ **do**
3:      SCCs = findSCCS(s, M, $F_A(f_A^U)$);
4:      **if** checkU(s, M, ϕ$_1$, ϕ$_2$, SCCs) = true **then**
5:         **return** true;
6:      **end if**
7:   **end for**
8:   **return** false;
9:**end function**
10:**function** checkU(s, M, ϕ$_1$, ϕ$_2$, SCCs):
11:   **for all** scc **in** SCCs **do**
12:      **for all** (s, M)'s successor (s', M') **in** scc **do**
13:         **if** ϕ$_1$ ∉ label(s') **or** CHECK(s', M', ϕ$_1$) = false **then**
14:            **return** false;
15:         **end if**
16:         **if** ϕ$_2$ ∈ label(s') **or**   CHECK(s', M', ϕ$_2$) = true **then**
17:            **break;**
18:         **end if**
19:      **end for**
20:   **end for**
21:   label(s) := label(s) ∪ {⟨⟨A⟩⟩(ϕ$_1$ U ϕ$_2$)};
22:   **return** true;
23:**end function**

**Algorithm 10** findSCCS(s, M, $F_A(f_A^{tU})$)

```
1: function findSCCS(s, M, F_A(f_A^tU)):
2:    stack := empty stack, SCCs := empty list of scc, v := new vertex(s, M), index := 0;
3:    strongConnect(v, SCCs, index, stack);
4:    return SCCs;
5: end function
6: function strongConnect(v, SCCs, index, stack):
7:    v.index := index;
8:    v.lowlink := index;
9:    index := index + 1;
10:   pushToStack(stack, v);
11:   v.onStack := true;
12:   for all (s', M')' in (s, M) →^act (s', M'), act ∈ F_A(f_A^tU)(s) do
13:      w := new vertex(s', M');
14:      if w.index is undefined then
15:         strongConnect(w, SCCs, index, stack);
16:         v.lowlink := min(v.lowlink, w.lowlink);
17:      else if w.onStack then
18:         v.lowlink := min(v.lowlink, w.index);
19:      end if
20:   end for
21:   if v.lowlink = v.index then
22:      scc := ∅;
23:      while true do
24:         w := popFromStack(stack);
25:         w.onStack := false;
26:         scc := scc ∪ {w};
27:         if w = v then
28:            break;
29:         end if
30:      end while
31:      SCCs := SCCs ∪ scc;
32:   end if
33: end function
```

The operators corresponding to Algorithms 7, 8, and 9, denoted as ⟨⟨A⟩⟩G, ⟨⟨A⟩⟩F, and ⟨⟨A⟩⟩U respectively, all leverage Algorithm 10 to reduce the time complexity by generating strongly connected components during the marking of process transitions. Algorithm 10 is implemented using the conventional Tarjan's algorithm[16-17]. In essence, it conducts a depth-first search (DFS) to record the traversal index, denoted as index, for each node, as well as the index value of the earliest vertex reachable from vertex v and its descendants via non-tree edges, which is referred to as the lowlink value. This algorithm enables the identification of strongly connected components in linear time. Consequently, the time complexity of Algorithms 7, 8, and 9 is O(N*(V+E)), where V+E represents the total number of vertices and edges across all connected components, and N is the number of partial strategies.

## 5. IMPLEMENT

We present a data structure for describing formulas termed the Formula Node Structure, which assumes a tree-like form. In this structure, the root node represents an operator, and the child nodes are also instances of the Formula Node Structure, denoting the data structure for the sub-formulas of the operator. This structure is simple and conducive to recursive calls on sub-formulas. For example, the Formula Node structure of ⟨⟨A⟩⟩($\phi_1$ U $\phi_2$) is shown in Fig.2.

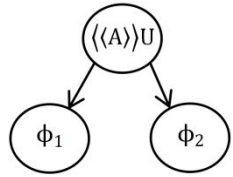

Fig.2. the Formula Node structure of $\langle\langle A\rangle\rangle(\phi_1 U \phi_2)$

The input consists of an EPC model and an ATLE formula $\phi$. Upon receiving a formula, the program recursively invokes sub-procedures to perform verification. Next, the implementation of the $C_A(\phi)$ and $\langle\langle A\rangle\rangle G(\phi)$ operators is mainly used as an example.

The Common Knowledge operator $C_A(\phi)$, denotes that a proposition $\phi$ is common knowledge among a group of agents A. This concept implies that not only does each agent in A know $\phi$, but they also know that others know $\phi$...,recursively, leading to an infinite regression of mutual knowledge. To implement $C_A(\phi)$ in the ATLE model, the CHECKC algorithm is designed to confirm whether the formula $\phi$ holds as common knowledge among the agents in A. The procedure begins by verifying that $\phi$ is satisfied for each agent in A at a given state. It then employs an iterative approach to verify the transitive propagation of knowledge between agents through epistemic accessibility relations $R_A^+$, which denote which states are accessible to each agent based on their knowledge. To ensure common knowledge, CHECKC recursively verifies that each agent within the set A maintains this knowledge across all related states.

The Global operator $\langle A\rangle\rangle G(\phi)$ asserts that there exists a strategy for a coalition of agents A that guarantees the formula $\phi$ remains true across all states along any path that the system might follow. The CHECKG algorithm implements this operator by exploring the state space iteratively, ensuring that $\phi$ is satisfied along every possible path dictated by the strategies available to agents in A. The algorithm leverages depth-first search (DFS) to find strongly connected components (SCCs) within the state space, which allows it to identify loops or recurring states that must consistently satisfy $\phi$. By locating these SCCs, the algorithm can verify that $\phi$ holds globally within each component, as any failure to satisfy $\phi$ at any state within an SCC would imply that $G\phi$ does not hold. The Strategy data structure defines the set of actions agents can perform, guiding the exploration of state transitions recorded in the Transition structure. Each strategy for agents in A generates different paths through the system's states, and the algorithm checks $\phi$ at each state along these paths until all possible strategies are evaluated or a counterexample is found.

## 6.CASE STUDY

### 6.1. Modeling

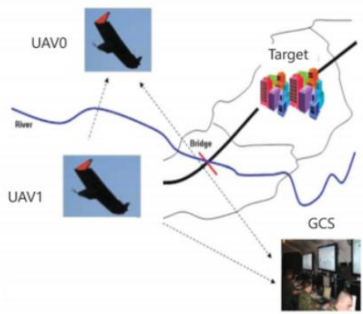

Fig.3. a system for multiple drones engagement

With the rapid development of UAV technology, drones are now widely used in military and civilian applications such as reconnaissance, strikes, and transport. In the military field, UAV swarms have become a key combat strategy, where multiple drones collaborate to perform missions like attacking targets or gathering intelligence (see Fig.3). However, UAV swarm operations face challenges, including coordinating actions to avoid conflicts, efficiently sharing information like target data, and ensuring mission safety and reliability against enemy interference. Advanced modeling and verification methods are crucial to address these issues. This study examines a cooperative system where three UAVs attack a single target under ground control. Using Epistemic Process Calculus (EPC) and Alternating-time Temporal Logic with Epistemic operators (ATLE), it explores modeling and verifying the system to ensure safety and reliability.

In order to precisely model the collaborative engagement system involving multiple UAVs, a set of atomic propositions is defined to represent the fundamental events and states of the system:

(1) $p_i$: $UAV_i$ has successfully struck the target.

(2) $q_i$: the $UAV_i$ receives the attack command.

The "com" and "attack" are two actions, where "com" represents the GCS sending cognition to all UAVs, and "attack" signifies a UAV targeting and attacking a certain objective while transmitting cognition.

Let $r = q_0 \wedge q_1 \wedge \neg p_1 \wedge \neg p_2$, the system is modelling as follows:

$$P_i = com(v).\overline{attack}<p_i>.P_i, \ 0 \le i \le 1$$

$$Q = \overline{com}<r>.(attack(v_1).attack(v_0) | attack(v_0).attack(v_1)).Q$$

The labeled process of the system is defined as follows:

$$M_0 = \{P_0\}_{UAV_0} | \{P_1\}_{UAV_1} | \{Q\}_{GCS}$$

The following presents the relevant symbolic definitions and notations:

(1) $Ag = \{UAV_0, UAV_1, GCS\}$.

(2) $S = \{s_0, s_1, s_2, s_3, s_4\}$.

(3) $ES = \{es_0, es_1, es_2\}$.

(4) $PS = \{\{P_0\}_{UAV0} = \{com(v).\overline{attack}<p_0>.P_0\}_{UAV0}, \{P_0'\}_{UAV0} = \{\overline{attack}<p_0>.P_0\}_{UAV0}, \{P_1\}_{UAV1} = \{com(v).\overline{attack}<p_1>.P_1\}_{UAV1}, \{P_1'\}_{UAV1} = \{\overline{attack}<p_1>.P_1\}_{UAV1}, \{Q\}_{GCS} = \{\overline{com}<r>.(attack(v_1).attack(v_0)|attack(v_0).attack(v_1)).Q\}_{GCS}, \{Q'\}_{GCS} = \{(attack(v_1).attack(v_0)|attack(v_0).attack(v_1)).Q\}_{GCS}\}$.

(5) $MS = \{M_0 = \{P_0\}_{UAV0}|\{P_1\}_{UAV1}|\{Q\}_{GCS}, M_1 = \{P_0'\}_{UAV1}|\{P_1'\}_{UAV1}|\{Q'\}_{GCS}, M_2 = \{P_0'\}_{UAV0}|\{P_1\}_{UAV1}|\{Q'\}_{GCS}, M_3 = \{P_0\}_{UAV0}|\{P_1'\}_{UAV1}|\{Q'\}_{GCS}\}$.

(6) $AP = \{p_0, p_1, q_0, q_1\}$.

(7) $acts = \{a_1 = \{\overline{com}<r>\}_{GCS}, a_2 = \{\overline{attack}<p_0>\}_{UAV0}, a_3 = \{\overline{attack}<p_1>\}_{UAV1}, \{\tau\}_{GCS,UAV_0}, \{\tau\}_{UAV_0,GCS}, \{\tau\}_{UAV_1,GCS}\}$.

(8) $K = \{(s_0, \{\tau\}_{GCS,UAV_0}, s_1), (s_1, \{\tau\}_{UAV_0,GCS}, s_2), (s_1, \{\tau\}_{UAV_1,GCS}, s_3), (s_2, \{\tau\}_{UAV_1,GCS}, s_4), (s_3, \{\tau\}_{UAV_0,GCS}, s_4), (s_4, \{\tau\}_{GCS,UAV_0,}, s_1)\}$.

(9) $\Delta = \{M_0 \xrightarrow{\{\tau\}_{GCS,UAV_0}} M_1, M_1 \xrightarrow{\{\tau\}_{UAV_0,GCS}} M_2, M_1 \xrightarrow{\{\tau\}_{UAV_1,GCS}} M_3\}$.

(10) $h_0(s_0) = es_0, h_1(s_0) = es_0, h_2(s_0) = es_0, h_0(s_1) = es_1, h_1(s_1) = es_1, h_2(s_1) = es_1, h_0(s_2) = es_2, h_1(s_2) = es_0, h_2(s_2) = es_0, h_0(s_3) = es_0, h_1(s_3) = es_2, h_2(s_3) = es_0, h_0(s_4) = es_2, h_1(s_4) = es_2, h_2(s_4) = es_2$.

(11) $T(s_0)=\emptyset$, $T(s_1)=\{q_0, q_1\}$, $T(s_2)=\{p_1, q_0, q_1\}$, $T(s_3)=\{p_0, q_0, q_1\}$, $T(s_4)=\{p_0, p_1, q_0, q_1\}$.

The aforementioned constitutes a set of pivotal information as delineated by the marked process. Herein, Ag represents the collective of agents within the system, S denotes the set of states, ES constitutes the collection of epistemic states, PS is the assemblage of processes for each agent, and MS encompasses the entire array of marked processes that can be generated subsequent to the transition of the system as a marked process. Additional definitions can be referenced in the formal specifications provided in preceding sections. In particular, $h_i(s_j)$ abstracts the epistemic situation of agent i to form the epistemic state es. For instance, $h_0(s_0) = es_0$ indicates that $UAV_0$ has not received an attack command. Different epistemic states under different states may indeed be identical. $s_0$ represents a global state that encompasses not only cognitive states but also various physical states of the agent, including positional information and network status. For instance, the position of an unmanned aerial vehicle being at 300 meters is a physical state, whereas the UAV's knowledge of its own position at 300 meters constitutes its cognitive state. The former pertains to positional information, while the latter is the focal point of this paper, representing the UAV's cognitive state. In addition, $es_1$ signifies that the agent is aware of the attack command. $es_2$ indicates that the agent is aware of the attack command and now that it initiated an attack. Following the establishment of these formal definitions, they can be employed as inputs, along with the definitions of relevant properties expressed using ATLE formulas, also considered as inputs. Subsequently, model checking algorithms can be utilized to ascertain the veracity of the specified properties.

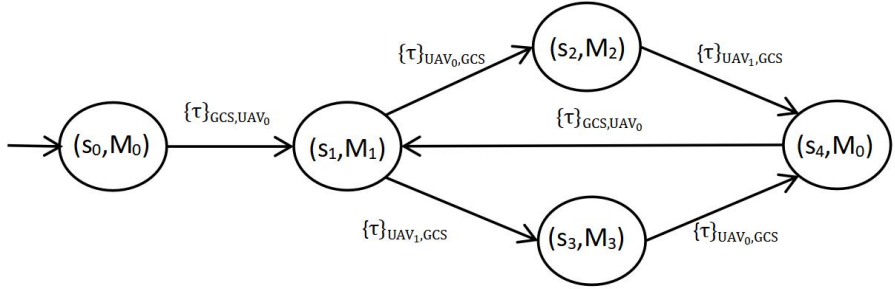

Fig.4. the transitions of the system

Refer to Fig.4. this is the transition model of the system. $(s_0, M_0)$ represents the initial preparatory state of the system, where the Ground Control Station (GCS) issues a command for the swarm of Unmanned Aerial Vehicles (UAVs) to initiate an attack on a specified target, transitioning the system into state $(s_1, M_1)$. S1 is the attack readiness state for two UAVs, each of which carries out an attack action, subsequently entering states $(s_2, M_2)$ and $(s_3, M_3)$, respectively. Following the attack actions, instructions are transmitted back to the GCS, moving the system into state $(s_4, M_0)$. Thereafter, the GCS can issue further attack commands, returning the system to state $(s_1, M_1)$, thereby establishing a recyclable chain of command.

## 6.2. Experiments Result

By substituting the formula into the algorithm in section 4.2 and the implement in section 5, the ATLE logic formula can be constructed and verified. The final evaluation results are show in the Table 3. Let $A = \{UAV_0, UAV_1, GCS\}$ and $B = \{UAV_0, UAV_1\}$ be two sets of agents.

Table 3. Some Logic Formulas Verification results.

| Logic Formula | Meaning | Result |
| --- | --- | --- |
| $(s_0, M_0) \vDash \langle\langle A \rangle\rangle G(C_A(q_1 \to q_0))$ | Within set A, there exists a strategy that ensures along this subsequent path, the cognition that "if $UAV_1$ receives an attack command, then $UAV_0$ also receives the command" becomes common knowledge. | true |
| $(s_0, M_0) \vDash \langle\langle A \rangle\rangle X(K_0(K_1(q_0)))$ | Within set A, there exists a strategy that in the next state, $UAV_0$ is aware that $UAV_1$ knows the knowledge "$UAV_0$ receive the command". | true |
| $(s_0, M_0) \vDash \langle\langle A \rangle\rangle F(E_A(p_0))$ | Within set A, there exists a strategy that finally, every agent know the knowledge "$UAV_0$ struck the target". | false |
| $(s_0, M_0) \vDash \langle\langle A \rangle\rangle F(D_A(q_0))$ | Within set A, there exists a strategy that $q_0$ will ultimately become a piece of distributed knowledge. | true |
| $(s_0, M_0) \vDash \langle\langle B \rangle\rangle F(D_A(q_0))$ | Within set B, there exists a strategy that, $q_0$ will ultimately become a piece of distributed knowledge. | false |

For the first property, since the GCS sends commands synchronously to all UAVs, it is evident that only in state $(s_1, M_1)$ are commands sent, and all UAVs receive both $q_0$ and $q_1$. Therefore, the property is true. For the second property, in the subsequent state $(s_1, M_1)$, since $h_1(s_1)=es_1$ and there are no epistemically equivalent states, it is clear that $K_1(q_0)$ holds. Similarly, when iteratively computing $K_0(K_1(q_0))$, the same reasoning applies. Thus, the property is validated as true only in state $(s_1, M_1)$. For the third property, after $UAV_0$ sends an attack command, it transitions to state $(s_2, M_2)$. In this state, both the GCS and $UAV_0$ become aware of $p_0$. Therefore, the property is false. For the fourth property, once the system enters the command loop phase, it continuously maintains the presence of $q_0$, which represents distributed knowledge. Hence, the property is true. For the fifth property, we have $f_B^U(s_0) = a_1$, $f_B^U(s_2) = \{\tau\}_{UAV_0,GCS}$, $f_B^U(s_3) = \{\tau\}_{UAV_1,GCS}$, $U = \{s_0, s_2, s_3\}$. Then we have

$$(s_0, M_0) \xrightarrow{\{\tau\}_{GCS,UAV_0}} (s_1, M_1), \{\tau\}_{GCS,UAV_0} \notin f_B^U(s_0)$$

So $out(f_{\{UAV_0,UAV_1\}}, (s_0, M_0)) = \emptyset$, the fifth property is false. Put simply, all strategies here restrict the actions of the GCS and there is no action from the GCS that can transition to the next state.

## 7. CONCLUSION AND FUTURE WORK

This paper presents a model checking framework for multi-agent systems based on Value-passing CCS, effectively addressing the complex epistemic and communication interactions inherent in such systems. The proposed ATLE logic and its model checking algorithm enable the verification of key properties, demonstrated through a case study on collaborative UAV mission. In the future, we are committed to investigating a broader spectrum of complex scenarios to further refine and validate the robustness of this framework.

## ACKNOWLEDGEMENTS

This work was supported by the Fundamental Research Funds for the Central Universities under Grant NJ2024030.